\documentclass[a4paper,11pt]{article}
\usepackage{pos}
\usepackage{multirow}

\title{The gamma-ray emission from 3HWC J1928+178}
 \ShortTitle{The gamma-ray emission from 3HWC J1928+178}

\author*{Armelle Jardin-Blicq}

\affiliation{National Astronomical Research Institute of Thailand (Public Organization), Don Kaeo, MaeRim, Chiang Mai 50180, Thailand}
\affiliation{Department of Physics, Faculty of Science, Chulalongkorn University, 254 Phayathai Road, Pathumwan, Bangkok 10330, Thailand}
\affiliation{Max-Planck Institute for Nuclear Physics, D-69117 Heidelberg, Germany}

\forColl{HAWC} 

\emailAdd{armelle@narit.or.th}

\abstract{The gamma-ray source 3HWC~J1928+178, discovered by HAWC, is coincident with the 82 kyr pulsar PSR~J1928+1746, located 4 kpc away. It has not been reported by any Imaging atmospheric Cherenkov Telescope (IACT), until the recent detection of emission from this region by H.E.S.S., using an analysis adapted to extended sources. No counterpart in GeV gamma-rays from Fermi-LAT data or in X-ray has been reported so far. In this contribution, I give the multiwavelength context of the region surrounding 3HWC~J1928+178 and present a multi-component model derived using the Multi-Mission Maximum Likelihood framework (3ML). I explore the possibility to model the gamma-ray emission of 3HWC~J1928+178 by an extended source with continuous diffuse emission. Together with the age of the pulsar and its extended nature, it may indicate a transition from a pulsar wind nebulae to a halo, where the electrons have started to cool and diffuse away from the source.}

\FullConference{37$^{\rm{th}}$ International Cosmic Ray Conference (ICRC 2021)\\
		July 12th -- 23rd, 2021\\
		Online -- Berlin, Germany}

\begin{document}
\maketitle

\section{Introduction}

\subsection{Astrophysical context}
3HWC~J1928+178 ($\ell=52.93$\textdegree, $b=0.2$\textdegree) is a gamma-ray source discovered by HAWC and reported in the 3HWC catalogue~\cite{3HWC_catalog} at the significance level of $\sim 15\sigma$. No Imaging Atmospheric Cherenkov Telescope (IACT) has detected it before the confirmation by the H.E.S.S. collaboration\cite{HAWC-HESS_GP}, using an analysis method adapted to extended sources. The origin of the observed very high energy gamma-ray emission is still unclear. It may be associated with the pulsar PSR~J1928+1746, discovered at radio wavelength~\cite{Discovery_PSRJ1928}, although neither the pulsar nor the pulsar wind nebulae (PWN) has been detected in the X-ray energy range. In the vicinity of 3HWC~J1928+178 is another HAWC source, 3HWC~J1930+188 ($\ell=54.03$\textdegree, $b=0.32$\textdegree), also detected by H.E.S.S.~\cite{HGPS} and VERITAS~\cite{VERITAS_J1930}, associated with the PWN in the supernova remnant SNR~G54.1+0.3. It hosts the pulsar PSR~1930+1852. Finally, another energetic pulsar is located nearby, PSR~J1932+1916, associated with the Fermi source 3FGL J1932.2+1916. The characteristics of the three pulsars are gathered in Table~\ref{pulsars}. 

\begin{table}[ht!]
 \caption{ Characteristics of the pulsars located in the vicinity of 3HWC J1928+178 taken from the ATNF catalogue ~\cite{ATNF}.
 }
 \centering
 \begin{tabular}{|c|c|c|c|}
\hline
    & PSR J1928+1746 & PSR J1930+1852 & PSR J1932+1916 \\
\hline
celestial coordinates (ra \textdegree, dec \textdegree) & (292.18,17.77) & (292.62,18.87) & (293.08,19.28) \\
\hline
galactic coordinates (l \textdegree, b \textdegree) & (52.93,0.11) & (54.1,0.26) & (54.67,0.09) \\
\hline
distance (kpc) & 4.3 & 6.2 & -  \\
\hline
age (kyr) & 82.6 & 2.9 & 35.4 \\ 
\hline
period (s) & 0.069 & 0.14 & 0.21 \\
\hline
spin down power (erg s$^{-1}$) & $1.6 \times 10^{36}$ & $1.2 \times 10^{37}$ & $4.1 \times 10^{35}$ \\
\hline
 \end{tabular}

  \label{pulsars}
\end{table}

\subsection{The HAWC observatory and HAWC Data}
The High Altitude Water Cherenkov (HAWC) gamma-ray observatory is located at a latitude of 19\textdegree N in Mexico, and at 4100 m in altitude. HAWC is composed of 300 water tanks instrumented with four photomultiplier tubes (PMT). When the secondary particles of an atmospheric air shower passes through HAWC, they produce Cherenkov light in the water tanks and each PMT records the time and amplitude of the signal. Combining the information of all the PMTs, we can build the footprint of the shower on the detector and reconstruct the parameters of the air shower. Each event is assigned to one of the 9 analysis bin depending on the fraction of the water tanks that has been hit. Bin definitions are given in~\cite{HAWC_crab}. For this analysis, 1523 days of HAWC data are used and events falling in the analysis bins 4 to 9 are selected. It corresponds to events triggering more than 25\% of the array, which gives an energy threshold of approximately 1~TeV. For this specific bin selection, the sources 3HWC~J1928+178 is refered to as HAWC~J1928+178 in~\cite{HAWC-HESS_GP}. The corresponding PSF of the instrument is $\sim$0.4\textdegree \ for bin 4 and decreases to less than 0.2\textdegree \ for bin~9. More information about the detector, the event reconstruction and the data analysis can be found in~\cite{HAWC_detector_ICRC2015} and~\cite{HAWC_crab}.

\section{Modeling the region around 3HWC J1928+178}
A multi-component fit based on a maximum likelihood approach is performed using the Multi-Mission Maximum Likelihood framework (3ML)~\citep{3ML} and the HAWC HAL\footnote{https://github.com/threeML/hawc\_hal} plugin. A model is defined for a 3.5\textdegree \ radius region around 3HWC~J1928+178, convolved with the instrument response and compared to the corresponding experimental data. Two models are considered, described in the following paragraphs.

\subsection{The four components model}
Following an iterative procedure, a first model is defined composed of 2 components: a point-like component and an extended symmetric Gaussian component at the location of the two HAWC sources 3HWC~J1930+188 and 3HWC~1928+178 respectively. This choice is motivated by the fact that 3HWC~J1930+188 was detected as a point-like source by HAWC, H.E.S.S.~\cite{HGPS} and VERITAS~\cite{VERITAS_J1930}, and by previous studies of 3HWC~J1928+178 that showed that it is likely extended~\cite{HAWC-HESS_GP}~\cite{ICRC_J1928}. 
A simple power law $\mbox{dN/dE} = \mbox{F}_0 (\mbox{E}/\mbox{E}_0)^{\Gamma} $ is assumed as energy spectrum for all components. The fit is performed with the position, size of the extended source, flux normalisation $\mbox{F}_0$ at $\mbox{E}_0=10$~TeV and spectral index $\Gamma$ as free parameters. 
For each model, a test statistic (TS) is computed that compares the likelihood that the region is represented by the model against the hypothesis that there is background fluctuations only: $ \mbox{TS} = 2~\mbox{ln}(\mathcal{L}(\mbox{model})/\mathcal{L}(\mbox{background}))$. According to Wilks' theorem, the quantity TS follows a $\chi^2$ distribution of N degrees of freedom, with N the difference in number of free parameters between the model and the background~\cite{Wilks}. 
If an excess is found in the residual maps, a new component is added at the location of the excess and the fit is performed again with the additional component. The procedure stops when the addition of a new component does not improve the fit by more than $\Delta\mbox{TS} = 25 $. 
The best model is found to be made of two point-like sources close to the location of the pulsars PSR~J1930+1852 and PSR~J1932+1916, and two extended sources represented by symmetric Gaussians. The first one is found to have a size $\sigma = 0.18$\textdegree \ (39\% containment) and is located at the position of 3HWC~J1928+178. The second one is very extended with a size $\sigma = 1.43$\textdegree \ and seems to cover the whole region, likely trying to account for some large scale gamma-ray emission, maybe diffuse emission from the galactic plane. The output parameters can be found in Table~\ref{model_parameters} and the model can be visualised on the top right panel of Figure~\ref{models}.
 
\subsection{The diffusion emission model}
\begin{figure}[ht!]
    \centering
    \includegraphics[width=1\linewidth]{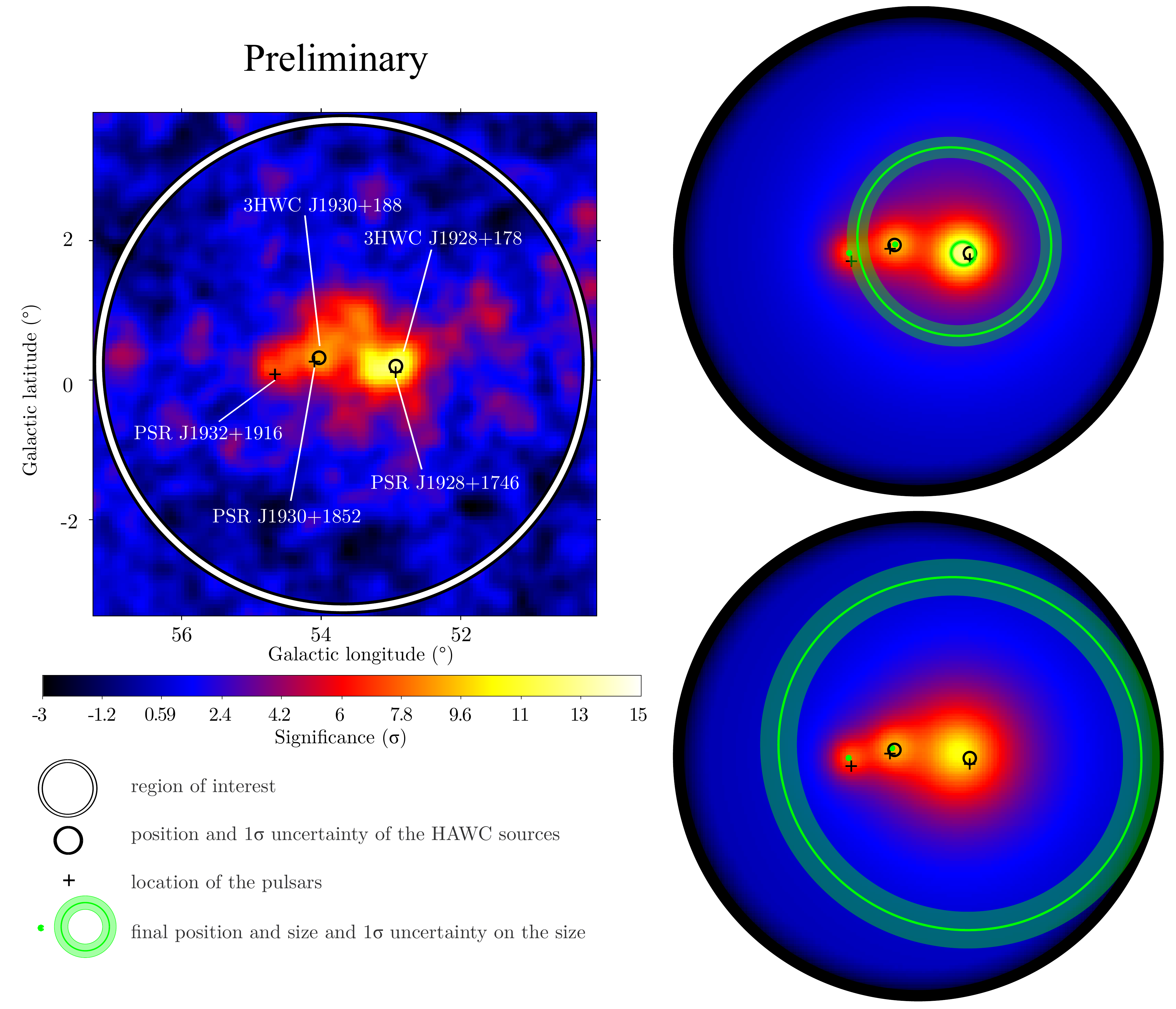}
    \caption{Left: HAWC significance map with the location of the HAWC sources and the location of the pulsars. The white circle is the region of interest of 3.5\textdegree \ radius. Right: Significance map of the 4 components model at the top and the diffusion model at the bottom, with the position and size of each fitted components. Each map is 3.5\textdegree \ in radius.
    }
    \label{models}
\end{figure}

Alternatively, a diffusion model is considered for 3HWC~J1928+178, assuming a continuous injection of electrons and positrons at the location of the remaining excess for 3HWC~J1928+178 after the first fit. 
Similar to the Geminga analysis~\cite{HAWC_geminga}, it is motivated by the fact that PSR~J1928+1746 is a rather old pulsar, that we don't see any X-ray counterpart and that the $\gamma$-ray emission seems extended. Indeed, the four components model shows that 3HWC~J1928+178 seems to be described by a superposition of a small Gaussian on top of a very wide one, which are both incorporated in this diffuse component for 3HWC~J1928+178. This model is thus composed of three components only: two point-like and one extended component with diffuse emission. As described in~\cite{HAWC_geminga}, the $\gamma$-ray flux $f_d$ for 3HWC~J1928+178, as a function of the distance from the source $d$, is approximately equal to: 
\begin{equation}
    f_{d} = \frac{1.22}{\pi^{3/2}r_d (d+0.06r_d)} \mbox{exp}\frac{-d^2}{r_d^2}
\end{equation}
where $r_d$ is the diffusion radius that is a free parameter fitted together with the position of the three components and their spectral parameter. The diffusion radius is found to be 2.68\textdegree. All the output parameters are gathered in Table~\ref{model_parameters} and the model can be visualised on the bottom right panel of Figure~\ref{models}.

 \begin{table}[ht!]
 \caption{ Output values from the fit for the components of the two different models representing 3HWC~J1928+178. 
 }
 \centering
 \begin{tabular}{|c|c|c|c|}
 \hline
                            & \multicolumn{2}{|c|}{4 components model}   & diffusion model \\
  \cline{1-4}
  hypothesis                 & Gaussian 1                 & Gaussian 2         & continuous injection   \\
  \cline{1-4}
  (ra \textdegree, dec \textdegree) & (292.15$_{\pm 0.03}$, 17.90{\tiny $\pm$0.04})  & (292.05{\tiny $\pm$0.15}, 18.10{\tiny $\pm$0.17})  & (292.10{\tiny $\pm$0.06}, 17.98{\tiny $\pm$ 0.02}) \\
  \cline{1-4}
  size ($\sigma$ or $r_d$)  (\textdegree)  &  0.18 {\tiny $\pm$ 0.03}   &  1.43 {\tiny $\pm$ 0.16}   & 2.68 {\tiny $\pm$ 0.27}  \\
  \cline{1-4}
  index                & -2.09 {\tiny $\pm$ 0.15}  & -2.60 {\tiny $\pm$ 0.08}  &  -2.58 {\tiny $\pm$ 0.05} \\
  \cline{1-4}
  flux$_{\mbox{\tiny10 TeV}}$ ($\times 10^{-15}$   & \multirow{2}*{4.2 {$_{\mbox{\tiny -1.1}}^{\mbox{\tiny +1.5}}$}} & \multirow{2}*{40 $_{\mbox{\tiny -4}}^{\mbox{\tiny +5}}$} & \multirow{2}*{47 $_{\mbox{\tiny -5}}^{\mbox{\tiny +5}}$}   \\
  ~TeV$^{-1}$~cm$^{-2}$~s$^{-1}$) & & & \\
  \cline{1-4}
  
 \end{tabular}

  \label{model_parameters}
\end{table}

\section{Comparison between the 2 models}
\begin{figure}[ht!]
    \centering
    \includegraphics[width=0.74\linewidth]{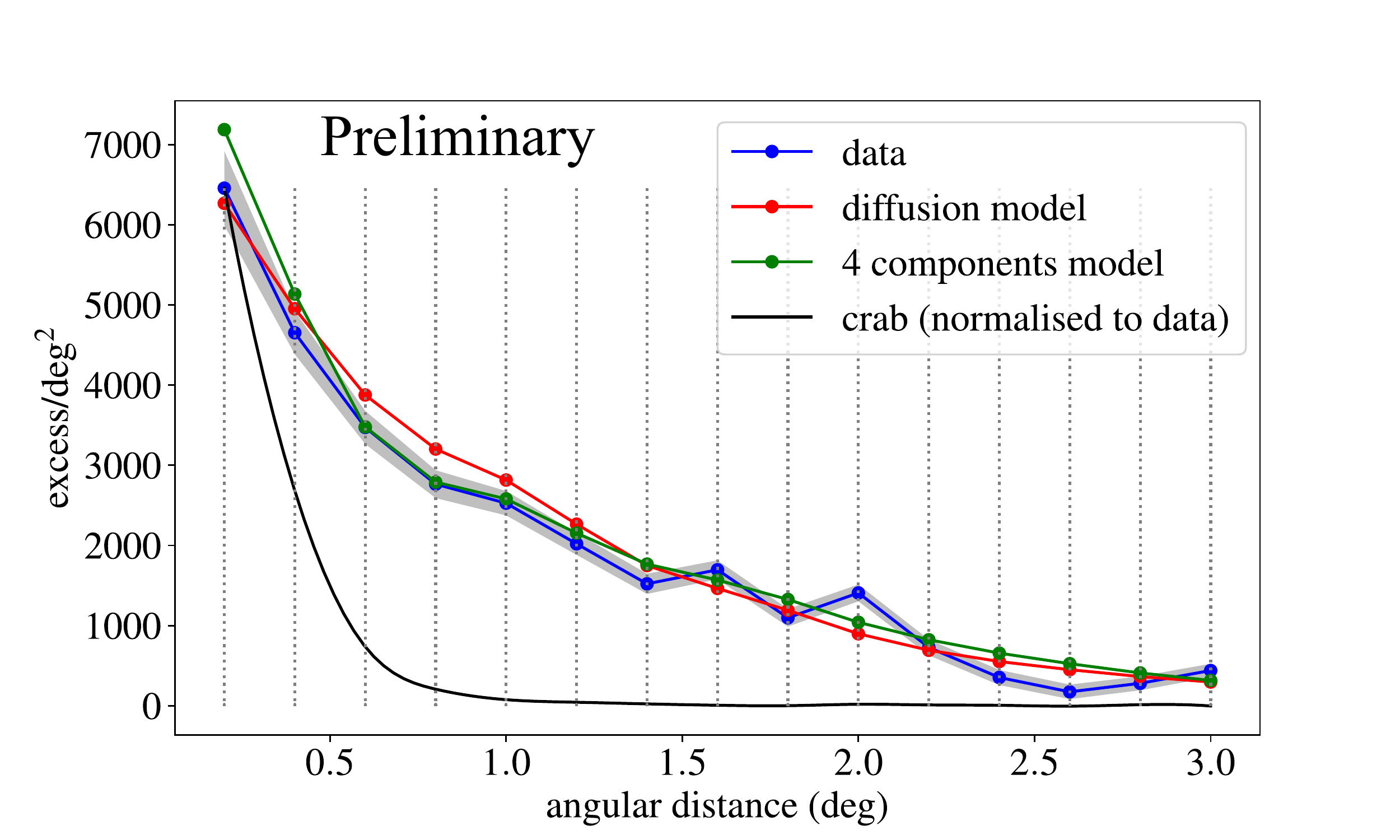}
    \caption{ Radial profile centered on J1928 in step of 0.2\textdegree, for the HAWC data (blue) with its standard deviation (grey band), the diffuse model (red) and the 4 components model (green). The profile of the models is plotted for the region of interest of 2\textdegree \ only. The profile of the crab nebula is also plotted (black) as a reference for the HAWC PSF. 
    }
    \label{model_profiles}
\end{figure}

The diffusion model is slightly worse by $\Delta$TS = 8 than the model with 4 components (2 point-like and 2 extended sources) but it has less degrees of freedom since it has only 3 components. 
To account for the difference in number of degrees of freedom we can look at the Bayesian information criterion number (BIC) for the two non-nested models used here~\cite{BIC}, given by this formula : $\mbox{BIC} = -2~\mbox{ln}(\mathcal{L}) + k\mbox{ln}(n)$ where $k$ is the number of free parameters and $n$ is the number of healpix pixels in the ROI. It seems that the diffusion model is largely preferred with a difference of $\Delta \mbox{BIC} = 45$.
Figure~\ref{model_profiles} shows the radial profiles centered on 3HWC~J1928+178 of the HAWC data, and for the two models. The profile of the Crab nebula is plotted as a reference for the HAWC PSF, showing that the emission is indeed extended. The profiles of both models follow closely the data.

\begin{figure}[h!]
    \centering
    \includegraphics[width=0.7\linewidth]{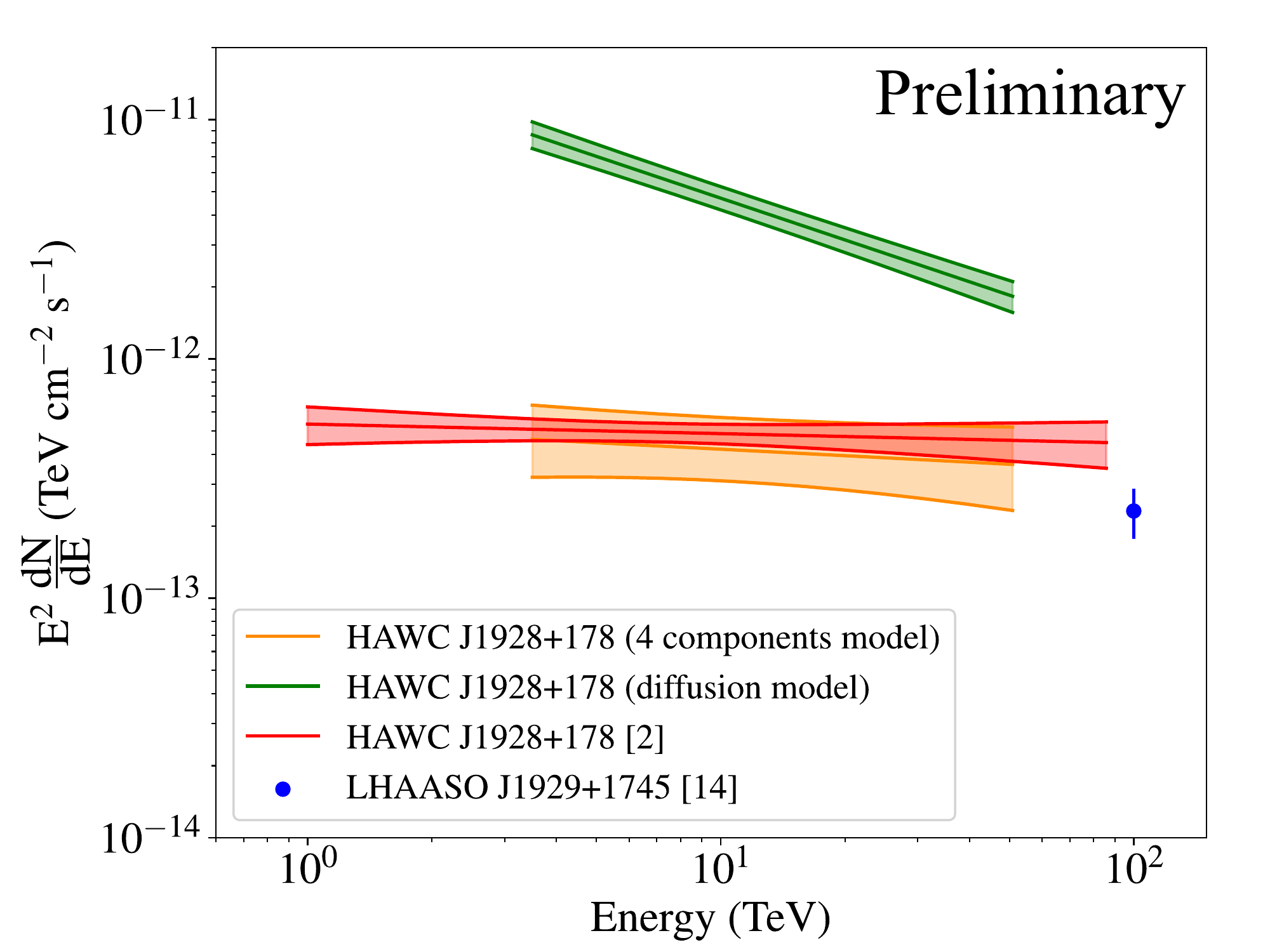}
    \caption{Spectral energy distribution of 3HWC~J1928+178 for the two models presented in this contribution, for the energy range 3.5~-~51~TeV, corresponding to the median energies of bin 4 and 9. The spectrum from the HAWC catalogue search for the same analysis bins~\cite{HAWC-HESS_GP} and the flux point from LHAASO are also plotted for comparison~\cite{LHAASO}. The shaded area correspond to the 1$\sigma$ statistical uncertainties.
    }
    \label{J1928_spec}
\end{figure}

The spectral energy distribution is plotted in Figure~\ref{J1928_spec} for the component representing the source 3HWC~J1928+178 for the two models. It is compared with the spectrum from the HAWC automatic source search for the same set of data~\cite{HAWC-HESS_GP}. The flux point from LHAASO~\cite{LHAASO} seems to be more compatible with the 4 components model, or imply a cut-off in the power low spectrum from the diffusion model.

\section{Discussion and conclusion}
The 4 components model and the diffusion model are very similar in the sense that they are in agreement with the hypotheses that 3HWC~J1928+178 is extended, and that the presence of a very large scale component is needed, either maybe due the galactic diffuse emission in one case, or to the diffusion of e$^\pm$ in the second case. The diffusion model, with 1 component less and a lower BIC, seemed favoured. It may indicate that 3HWC~J1928+178 is in a transition phase to a TeV gamma-ray halo: the pulsar being rather old, the TeV gamma-ray emission being extended, and the fact that no PWN has been detected in X-ray are favouring this hypothesis. 
However, the fitted diffusion radius $r_d$ is found to be 2.68\textdegree, \textit{i.e} $\sim400$~pc in diameter, since PSR~J1928+1746 is located 4.3~kpc away. It is huge compared to Geminga which has a diffusion radius of 5.5\textdegree \ but is only 250 pc away from us, giving a size of 23 pc in diameter. This may disfavour this model. Moreover, the spectrum for 3HWC~J1928+178 from the 4 components model is in better agreement with the measurement from LHAASO. A deeper study at the highest energy, using the energy estimators developed by the HAWC collaboration~\cite{HAWC_crab_100TeV} instead of the analysis bins would help to clarify the presence of a cut-off. 


\section*{Acknowledgments}
We acknowledge the support from: the US National Science Foundation (NSF); the US Department of Energy Office of High-Energy Physics; the Laboratory Directed Research and Development (LDRD) program of Los Alamos National Laboratory; Consejo Nacional de Ciencia y Tecnolog\'ia (CONACyT), M\'exico, grants 271051, 232656, 260378, 179588, 254964, 258865, 243290, 132197, A1-S-46288, A1-S-22784, c\'atedras 873, 1563, 341, 323, Red HAWC, M\'exico; DGAPA-UNAM grants IG101320, IN111716-3, IN111419, IA102019, IN110621, IN110521; VIEP-BUAP; PIFI 2012, 2013, PROFOCIE 2014, 2015; the University of Wisconsin Alumni Research Foundation; the Institute of Geophysics, Planetary Physics, and Signatures at Los Alamos National Laboratory; Polish Science Centre grant, DEC-2017/27/B/ST9/02272; Coordinaci\'on de la Investigaci\'on Cient\'ifica de la Universidad Michoacana; Royal Society - Newton Advanced Fellowship 180385; Generalitat Valenciana, grant CIDEGENT/2018/034; Chulalongkorn University’s CUniverse (CUAASC) grant; Coordinaci\'on General Acad\'emica e Innovaci\'on (CGAI-UdeG), PRODEP-SEP UDG-CA-499; Institute of Cosmic Ray Research (ICRR), University of Tokyo, H.F. acknowledges support by NASA under award number 80GSFC21M0002. We also acknowledge the significant contributions over many years of Stefan Westerhoff, Gaurang Yodh and Arnulfo Zepeda Dominguez, all deceased members of the HAWC collaboration. Thanks to Scott Delay, Luciano D\'iaz and Eduardo Murrieta for technical support.

\bibliographystyle{JHEP}
\bibliography{biblio.bib}

\clearpage
\section*{Full Authors List: \Coll\ Collaboration}
\scriptsize
\noindent
A.U. Abeysekara$^{48}$,
A. Albert$^{21}$,
R. Alfaro$^{14}$,
C. Alvarez$^{41}$,
J.D. \'Alvarez$^{40}$,
J.R. Angeles Camacho$^{14}$,
J.C. Arteaga-Vel\'azquez$^{40}$,
K. P. Arunbabu$^{17}$,
D. Avila Rojas$^{14}$,
H.A. Ayala Solares$^{28}$
% ...

\noindent
% ...
$^{14}$Instituto de F\'isica, Universidad Nacional Aut\'onoma de M\'exico, Ciudad de M\'exico, M\'exico,
% ...
$^{17}$Instituto de Geof\'isica, Universidad Nacional Aut\'onoma de M\'exico, Ciudad de M\'exico, M\'exico,
% ...
$^{21}$Physics Division, Los Alamos National Laboratory, Los Alamos, NM, USA,
% ...
$^{28}$Department of Physics, Pennsylvania State University, University Park, PA, USA,
% ...
$^{40}$Instituto de F\'isica y Matem\'aticas, Universidad Michoacana de San Nicol\'as de Hidalgo, Morelia, Michoac\'an, M\'exico,
$^{41}$FCFM-MCTP, Universidad Aut\'onoma de Chiapas, Tuxtla Guti\'errez, Chiapas, M\'exico,
% ...
$^{48}$Department of Physics and Astronomy, University of Utah, Salt Lake City, UT, USA,
% ...
%

\end{document}